\documentclass[11pt]{article}

\usepackage{ifthen}
\usepackage{graphicx}

\usepackage{amssymb}
\usepackage{amsmath}

\begin{document}

\title{Earthquake Forecasting Using Hidden Markov Models}
\author{Daniel W. Chambers\footnote{Mathematics Department, Boston College, chambers@bc.edu} , Jenny A. Baglivo\footnote{Mathematics Department, Boston College, baglivo@bc.edu} , John E. Ebel\footnote{Weston Observatory, Department of Earth and Environmental Sciences, Boston College, ebel@bc.edu} , Alan L. Kafka\footnote{Weston Observatory, Department of Earth and Environmental Sciences, Boston College, kafka@bc.edu}}

\maketitle
\thispagestyle{empty}

\begin{abstract}
This paper develops a novel method, based on hidden Markov models, to forecast earthquakes and applies the method to mainshock seismic activity in southern California and 
western Nevada. The forecasts are of the probability of a mainshock within  one, five, and ten days in the entire study region or in specific subregions and are based on the observations available at the forecast time, namely the interevent times and locations of the previous mainshocks and the elapsed time since the most recent one. Hidden Markov models have been applied to many problems, including earthquake classiÞcation; this is the first application to earthquake forecasting. \end{abstract}

{\bf Short title:} Forecasting with hidden Markov models


Corresponding author: Daniel W. Chambers, chambers@bc.edu, fax 617-552-3789

\vfill
\eject 

\section{Introduction}

In this paper, we develop a general method for forecasting future observations, given past observations, when the observations arise from a hidden Markov model. We then formulate a hidden Markov model for mainshock seismic activity in southern California and western Nevada and use the model to devise waiting time forecasts for this region, and to make (retroactively) daily forecasts for the period between June 1982 and December 2008 (more than nine thousand forecasts). Model parameters were estimated using data in the region from February 1932 through July 1964. 

For each forecast we used the data known at the forecast time to estimate the probability of a mainshock earthquake of magnitude at least 4.0 within 24 hours in the study region.  Five and ten day forecasts were also computed. The forecasts were then compared to the observed occurrence of earthquakes. The goal was to develop a method to identify times when there is a heightened probability of a mainshock in the region. We also extended the model and forecasts to include location as well as interevent time.

A hidden Markov model (HMM) consists of two random sequences $X_1, X_2, \ldots$ and $Y_1,Y_2, \ldots$ The first is called  the state sequence and the second the observation sequence. The state sequence is unknown (hidden), while the observation sequence is seen.
The state sequence is a Markov chain. At each time $t$, the observation $Y_t$ depends on the state of the system- that is, the value of $X_t$. The distribution of $Y_t$ is determined by a probability density function whose parameters depend on  the value of $X_t$. 

Hidden Markov models were developed by Baum and colleagues (Baum et al., 1966, 1970, 1972) and saw early use in speech recognition (Rabiner, 1989), but have been used for a variety of applications, including modeling ion channels (Fredkin and Rice, 1992a, 1992b) and modeling genetic sequences (Durbin et al., 1998; Clote and Backhofen, 2000). They have recently begun appearing in seismological studies (Granat and Donnellan, 2002; Beyreuther et al., 2008; Chambers et al., 2003; Ebel et al., 2007;  Wu, 2010); this paper fills in the details of the forecasting methodology outlined in Ebel et al. (2007) and applies it for the first time. Tsapanos (2001) used a Markov model with states corresponding to six geographic regions, to forecast the locations of large earthquakes in South America. 

The states in our first application correspond to a tendency for a shorter or longer waiting time from the previous mainshock; location is included in the second application. Given a state, the interevent time distribution was taken to be exponential, with a mean determined by that state. 

In Section 2, we  review HMM's, including algorithms for estimating model parameters. In Section 3, we develop the methodology for constructing post-event and scheduled forecasts of the probability of an earthquake within a given time period in the study region, as well as confidence intervals for the waiting time. In Section 2.2 and Section 3, we assume state-specific exponential distributions for the interevent times, but this could be generalized to other types of distributions. Section 4 applies the methodology to the data from southern California and western Nevada. Section 5 extends the model and forecasts to include location as well as interevent time and Section 6 is a summary and discussion. Details on the derivations are contained in the appendix.

\section{Hidden Markov Models}
 
We assume that $S$, the set of states, is finite and $(X_1, X_2, \ldots)$ is a Markov chain with transition probability matrix $A=\{a_{rs}, \, r,s \in S \}.$  That is, the distribution of $X_t$, conditional on previous values of the process, depends only on $X_{t-1}$ and the one-step transition probabilities are given by $A$:  $P(X_t=s | X_{t-1}=s_{t-1}, \ldots , X_1=s_1)=P(X_t=s | X_{t-1}=s_{t-1})$ and $P(X_t=s | X_{t-1}=r)=a_{rs}$.  

The observations  can be discrete or continuous random variables. We assume the distribution of the observation $Y_t$ at time $t$, knowing the system is in state $s$, is determined by a probability density function $p_s(y)=P(Y_t=y | X_t=s)$. We use this notation for continuous as well as discrete observation variables; in the continuous case, the density is the derivative of the cumulative distribution function: $p_s(y)=\frac{d}{dy} P(Y_t \le y |X_t=s)$.  We further assume that, conditional on $X_t$, $Y_t$ is independent of the state and observation variables at all other times- that is, 
\begin{equation}
P(Y_t=y | X_t=s, X_{t'}, Y_{t''})=P(Y_t=y | X_t=s) {\rm~for~ all~} t',t'' \ne t.
\end{equation}
An induction argument and summing over all values of intermediate states easily establishes that, knowing the state at time $t$, the distribution of $Y_{t+k}$ is independent of the state and observation variables at times preceding $t$:
\begin{equation}
P(Y_{t+k} =y | X_t=s, X_{t'}, Y_{t''})=P(Y_{t+k}=y | X_t=s) {\rm~for~all~} k \ge 0 {\rm ~and~} t', t'' \le t.
\end{equation}
Let $\pi = (\pi_s, \, s \in S)$ denote the initial distribution of the Markov chain: $\pi_s=P(X_1=s)$. Let $\Theta$ be the collection of parameters in the HMM; these consist of the transition probabilities, the initial distribution vector, and the state-specific distribution parameters.  In the following, we assume the state and observation sequences are of length $L$. For observations $y_1, \ldots , y_t$, we abbreviate $P(Y_1=y_1, \ldots, Y_t=y_t ) ~{\rm by}~ P(y_1, \ldots, y_t ).$ Finally, the complete set of observations $y_1, \ldots, y_L$ will be denoted $\mathcal{O}$ to simplify notation. Thus, $P(\mathcal{O} )$ is the likelihood of the observations under the HMM with parameter set $\Theta$.

\subsection{Known parameters}
 
 Our estimation and forecasting techniques utilize the so-called forward and backward variables, $f_s(t)$ and $b_s(t)$, respectively. At time $t$ and state $s$, the forward variable $f_s(t)$ represents the joint probability of the observations up to time $t$ and being in state $s$ at that time: $f_s(t)=P(y_1, \ldots, y_t, X_t=s )$. The backward variable at time $t$ and state $s$ gives the joint probability of the future observations conditional on being in state $s$ at time $t$: $b_s(t)=P(y_{t+1},...,y_L | X_t=s).$
 
 Assume the collection of parameters $\Theta$ is known; then a well-known method (Rabiner, 1989) exists for  efficiently calculating the forward and backward variables for each time $t=1, \ldots L$ and state $s \in S$ via the forward and backward equations, a set of initializations and recursions. Namely, for all states $s$ and times $1 \le t \le L-1$,
\begin{equation}
 f_s(1)=p_s(y_1)\pi_s ,\ \ \ \ \ \  f_s(t+1)=\sum_{r \in \mathcal{S}}f_r(t)a_{rs}p_s(y_{t+1}),
\end{equation}
\begin{equation}
b_s(L)=1, \ \ \ \ \ \ \ \ \ \ \  b_s(t)=\sum_{r \in \mathcal{S}} p_r(y_{t+1})b_r(t+1)a_{sr}.
\end{equation}
As a first application, we note that the likelihood of the observations up to time $t$, $y_1, \ldots , y_t$,  is easily computed from the forward variables: $P(y_1, \ldots , y_t )=\sum_{s \in S}P(X_t=s,y_1, \ldots , y_t )$, so that 
\begin{equation}
P(y_1, \ldots , y_t )=\sum_{s \in S}f_s(t), ~ t=1, \ldots , L.
\end{equation}
In particular,
\begin{equation}
P(\mathcal{O} )=\sum_{s \in S}f_s(L).
\end{equation}

\subsection{Estimating parameters}

For parameter estimation and forecast evaluation, we used separate portions of a catalog of 1300 earthquakes that occurred in southern California and western Nevada, from February, 1932 through December, 2008. This is the collection of all magnitude 4 or larger mainshock earthquakes; foreshocks and aftershocks were removed from the complete catalog using the method of Gardner and Knopoff (1974). See Figure 1 for the study region and the earthquake locations.  Throughout this paper, earthquake will always refer to a magnitude 4 or larger mainshock.
 
[INSERT FIGURE 1 ABOUT HERE]

In our formulation of the HMM, time corresponds to earthquakes in the study region. That is, time $t$ refers to the $t^{th}$ earthquake. The interevent time between two earthquakes is measured in days. 

We consider a two state hidden Markov model for mainshocks in the entire study region. The two states correspond to an increased likelihood for a shorter or longer interevent time between earthquakes.

For the state-specific observation distributions,  each observation $y$ is the interevent time in days since the previous earthquake.  Given a state $s \in \mathcal{S}$, we assume an exponential distribution for interevent time, with mean $\lambda_s$.  Thus, the state-specific densities are: for  $s \in \mathcal{S}$ and $y>0$, 
\begin{equation} 
p_s(y)=(1/\lambda_s)e^{-y/\lambda_s} 
\end{equation}.

Thus, the parameter set consists of 
\begin{equation}
\Theta=\{a_{rs},  \, \pi_s, \, \lambda_s,  ~{\rm for~} r,s \in \mathcal{S} \}
\end{equation}

The standard method  for estimating the parameter set $\Theta$ for an HMM is called the Baum-Welch algorithm (Rabiner, 1989). It is a special case of the Expectation-Maximization method. (See Granat and Donnellan, 2002 for a good explanation of the connection.)  Details are given in the appendix. 

\section{Forecasts}

We are interested in forecasting the probability of an earthquake in the entire study region within the next day, next week, or any given time interval. There are two kinds of forecasts of interest. The first  type of forecast, which we call a post-event forecast, is issued immediately after an earthquake and the second type, which we call a scheduled forecast, is issued on a scheduled basis, say daily or weekly.  In Section 3.1, we present the post-event forecast methodology and in  Section 3.2 we handle scheduled forecasts. In Section 3.3 we derive moments for waiting times, conditional on the available history at the forecast time. In Section 3.4, we show that the  HMM's satisfy the waiting time paradox of Davis et al. (1989) and Sornette and Knopoff (1997).

\subsection{Post-event forecasts}

In this section, we first give the density function for waiting time until the next earthquake, made immediately after an earthquake. The derivation is given in the appendix. We then integrate this density to find the post-event forecast probability of an earthquake within a specified time interval.

Recall that in our application the observation $Y_t$ corresponding to earthquake $t$ is  the time in days since earthquake $t-1.$ The  density of the waiting time until the next earthquake, given the seismic history available just after an earthquake is
\begin{equation}
P(Y_{t+1}=y \, | \, y_1, \ldots y_t)=\sum_{s \in \mathcal{S}}c_{st}(1/\lambda_s)e^{-y/\lambda_s}.
\end{equation}
where 
\begin{equation}
c_{st}=\sum_{r \in \mathcal{S}}a_{rs} f_r(t)/P(y_1, \ldots y_t).
\end{equation}

Furthermore, as shown in the appendix,
\begin{equation}
c_{st}= P(X_{t+1}=s \, | \, y_1, \ldots , y_t). 
\end{equation}

Thus, (9)-(11) show that the forecast density of the next observation, conditional on the data available immediately after an earthquake, is a mixture of the state-specific densities, in which the weight for the exponential density $(1/\lambda_s)e^{-y/\lambda_s}$ is (a) calculable and (b) interpretable as the conditional probability the system is in that state $s$ given the data available at the forecast time.

Our forecast probability of an earthquake within $N$ days, made immediately after an earthquake and using the data available at  that time  is simply the integral of (9) from $0$ to $N$:
\begin{equation}
P(Y_{t+1} \le N \, | \, y_1, \ldots , y_t)=\sum_{s\in \mathcal{S}} c_{st}(1-e^{-N/\lambda_s}).
\end{equation} 
Note that this is a dynamic process: different forecast times have different seismic histories and (by (10)) give different forecast probabilities of an earthquake in a given time period. This is in marked contrast to the standard Poisson model which yields the same exponential density for all interevent times, regardless of seismic history.

\subsection{Scheduled Forecasts}

Next, we  extend our method to scheduled forecasts, say weekly or daily, of the probability of an earthquake within a given time period in the region. Now what is known at the time of the forecast is the set of observed interevent times as well as the elapsed time since the previous earthquake.

Suppose that $t$ earthquakes have occurred up to the time that the forecast is made and $w$ days have elapsed since the most recent earthquake. The available data at forecast time is then  $y_1, \ldots , y_t$ and the event $Y_{t+1} \ge w$. The remaining waiting time until the next event is $Y_{t+1}-w$. As shown in the appendix, the forecast density of the remaining waiting time, given the available data, is 

\begin{equation}
P(Y_{t+1}-w=y \, | \, y_1, \ldots, y_t, Y_{t+1} \ge w) = \sum_{s \in \mathcal{S}}d_{st}(w) (1/\lambda_s)e^{-y/\lambda_s} ,\\
\end{equation}
where 
\begin{equation}
d_{st}(w)=\frac{c_{st}e^{-w/\lambda_s}}{\sum_{r \in \mathcal{S}}c_{rt}e^{-w/\lambda_r}}.
\end{equation}

It is easily shown that $d_{st}(w)=P(X_{t+1}=s \, | \, y_1, \ldots, y_t, Y_{t+1} \ge w)$, so that the forecast density is again a mixture of the densities $p_s(y)$ with the weight for  $p_s(y)$ being the conditional probability of being in state $s$ at the next earthquake, given the data available at the forecast time. 

Paralleling (12), the scheduled forecast probability of an earthquake within $N$ additional days is obtained by taking the $t$ interevent times for the earthquakes that have occurred by that time, computing the elapsed time $w$ since the most recent earthquake, calculating (10) and (14), and integrating (13)  from $0$ to $N$, to give:
\begin{equation}
P(Y_{t+1}-w \le N  \, | \, y_1, \ldots, y_t, Y_{t+1} \ge w) = \sum_{s \in \mathcal{S}}d_{st}(w) (1-e^{-N/\lambda_s}) .\\
\end{equation}

Post-event forecasts are scheduled forecasts with an elapsed waiting time of $w=0$. Note that, by (14) and (42), $d_{st}(0)=c_{st}$ and (15) simplifies to (12) when $w=0$.

\subsection{ Moments for waiting time}

In addition to generating forecast probabilities, we can use these densities to find post-event and scheduled means and variances of waiting time, conditional on the seismic history. For example, the mean and variance of the remaining waiting time given that $t$ earthquakes have occurred and $w$ days have elapsed already are
\begin{equation}
E(Y_{t+1}-w \, | \, y_1, \ldots, y_t, Y_{t+1} \ge w)= \sum_{s \in \mathcal{S}}d_{st}(w) \lambda_s {\rm ~and~}
\end{equation}
\begin{equation}
V(Y_{t+1}-w \, | \, y_1, \ldots, y_t, Y_{t+1} \ge w)= \sum_{s \in \mathcal{S}}d_{st}(w) \lambda_s^2,\end{equation}
respectively.  
\subsection{Waiting time paradox}

Davis et al. (1989) and Sornette and Knopoff (1997) investigated the waiting time paradox, asking the question: can it be that the longer it has been since the last earthquake, the longer the expected time until the next? In our context, the question asks if $E(Y_{t+1}-w| y_1, \ldots y_t, Y_{t+1} \ge w)$ is an increasing function of $w.$ The answer depends on the assumed distribution of the intervent times. Sornette and Knopoff considered a variety of distributions, including the Periodic, Uniform, Gaussian, Semi-Gaussian, Lognormal, Weibull, Power Law, and Truncated Power Law.  For some distributions the expected waiting time was shown to be an increasing function of elapsed time $w$ over all values of $w$, for others a decreasing function over all values of $w$, in some cases increasing for ``small" values of $w$ and decreasing for ``large" values, and in others the opposite. In all of these, there was no HMM framework, no hidden states, and no state-specific distributions: $Y_1, Y_2 \ldots$ were considered independent and identically distributed. 

The waiting time paradox remains true for hidden Markov models with state-specific exponential distributions.  That is, the mean waiting time for the next earthquake increases as the elapsed time $w$ increases, for all $w$.  Specifically, as $w$ increases more weight is given to the larger state-specific expectations in the formula for the expected waiting times.  See the appendix for details. 

\section{Application to California/Nevada Data}

To illustrate our method we first developed a two-state hidden Markov model for seismicity in the study region.  In this section we give the parameter estimates based on the 601 earthquakes observed from February, 1932 through July, 1964 and then summarize the results of 9693 daily forecasts retrospectively made during the period June, 1982 through December, 2008.

The time periods for this illustration were chosen so that the assumption of stationarity of interevent time distributions was clearly met.  See Figure 2 for the complete distribution of interevent times for our catalog.

[INSERT FIGURE 2 ABOUT HERE]

For our two state model, State 1 corresponded to a tendency for a shorter interevent time and State 2 to a longer. Observations were interevent times and, given state $s$, were assumed to have an exponential distribution with mean $\lambda_s$. The parameter estimates were obtained by starting with uniform transition probabilities and a grid of exponential means $\{(i,j): i=1,4,7,10; \, j=10, 20, 30, 40, 50, 60, 70\}$. The Baum-Welch algorithm was run for approximately 100 iterations and the starting values that led to the highest likelihood were then used as the initial means and the algorithm was run until successive iterations led to a maximum difference in mean estimates and transition probability estimates of .000001.   The resulting estimates for $\Theta$ were:
\begin{equation}
(\hat \lambda_1,\hat  \lambda_2)=(1.4, \, 21.1), \quad (\hat  \pi_1,\hat  \pi_2)=(0,1), \quad \hat  A=\begin{pmatrix}.446 & .554 \cr .040 & .960\end{pmatrix}.
\end{equation}

Beginning on June 16, 1982, we (retrospectively) made daily forecasts of the probability of an earthquake within one day anywhere in the study region, based on the data available at the forecast time; these data consisted of the interevent times of the earthquakes seen so far and the elapsed time $w$ since the most recent earthquake. (For the observation history for the first forecast, we used the data  from the previous 30 earthquakes.) The desired forecast probability  followed immediately from (15) with $N=1$ day, where $d_{st}(w)$ was calculated from (14), (10), (3), and (5). This was repeated daily until December  28, 2008, resulting in 9693 forecasts.

The forecast probabilities ranged from a low of .0463 to a high of .2072. A sorted list of the forecasts is shown in Figure 3. To evaluate the forecasts, we split the 9693 days into two groups, those whose forecasts were in a ``low'' range and those with forecasts in a ``high'' range. Since most of the forecasts clustered near the low end, in this and the following analyses we took the lowest 9000 forecasts to define the low range and the remaining 693 to define the high range. The cutoff for this division is given by the horizontal line in Figure 3. Within each group, we then found the proportion of days on which an earthquake did occur within 24 hours and compared it to the interval, mean, and median forecast probabilities for that group.

[INSERT FIGURE 3 ABOUT HERE]

For the 9000 forecasts in the low range (.0463, .0558), the mean forecast was .0470 and the median was .0463; an earthquake occured within one day 444 times, for an observed proportion of .0493. (That is, on 9000 days between 1982 and 2008, the probability of an earthquake in the study area within 24 hours was forecast to be between 4.6\% and 5.6\%. Such an event did occur 4.9\% of the time.)  The observed proportion was in the forecast range and very close to the mean and median forecasts in that group. The 693  forecasts in the high range  (.0558, .2072) had a mean of .0672 and a median of .0616; the observed proportion was .0592, again very close to the mean and median forecasts in that group. The one day results are given in Table 1 and Figure 4.

[INSERT TABLE 1 ABOUT HERE]

[INSERT FIGURE 4 ABOUT HERE]

Each day we also forecast the probability of an earthquake within five days and ten days, using (15) with $N=5$ and $N=10$, respectively. For the five day forecasts, the low range had a mean of .2121 and a median of .2110, with an observed proportion of .2203. The high range mean was .2453, the median was .2361, and the observed proportion was .2626. Again, the forecasts appeared to do a good job. The results are given in Table 2 and Figure 5.

[INSERT TABLE 2 ABOUT HERE]

[INSERT FIGURE 5 ABOUT HERE]

For the ten day forecasts, the low range had a mean of .3784 and median of .3775, with an observed proportion of .3921. The high range mean was .4055, the median was .3980, and the observed proportion was .4271. In both cases, the forecasts seemed effective. The results are given in Table 3 and Figure 6.

[INSERT TABLE 3 ABOUT HERE]

[INSERT FIGURE 6 ABOUT HERE]

It is interesting to compare the forecast probabilities with the actual dates of earthquake events.  
One day, five day, and ten day forecasts are shown as time series in Figure 7, and the event dates are represented by vertical line segments along the bottom axis.  Peak forecast probabilities generally correspond to subsequent clustering of events.  

[INSERT FIGURE 7 ABOUT HERE]

The explanation for this lies in the note following (14), that the scheduled forecast density on a given day is a mixture of the two exponential densities, with weights given by the conditional probabilities of being in the two states, given the seismic history known at that point. If a number of earthquakes have occurred in a short amount of time, the conditional probability will be higher for State 1 than State 2, favoring the exponential with the shorter mean and hence leading to a higher forecast probability of an earthquake within one day. The top row of lines in Figure 5 mark the dates of the 500 earthquakes in the forecasting period. As can be seen, when a number of earthquakes occur in close proximity, this is reflected below by the subsequent appearance of high probability forecasts. 

The forecast methodology is effective for longer term forecasts as well, though it is less informative since the range of forecasts decreases. We derived thirty and one hundred day forecasts as above ($N=30, \, 100)$. For the former, the range of forecasts was $(.759,.842)$; the lower and higher groups had medians of $.759$ and $.767$ and observed proportions of $.786$ and $.794$, respectively. For the hundred day forecasts, the range was $(.9913,.9915)$ and the observed proportions were both $1$. The reason for this convergence of forecasts to $1$ is that in (15) the term $(1-e^{-N/\lambda_s})$ for large $N$ is essentially $1$ for both parameters $\lambda_1$ and $\lambda_2$ and, by (14), $\sum_{s \in S}d_{st}(w)=1$.

\section{Location and time forecasts}

We also developed a  four state HMM that incorporated location as well as time to derive forecasts of the probability of an earthquake within a given time period and in a given region.   In this model, called the  East/West model, region $1$ was East (quadrants 1 and 2) and region $2$ was West (quadrants 3 and 4; see Figure 1.) State ($r_1,r_2$) thus refers to a tendency for a short ($r_1=1$) or long ($r_1=2$) interevent time and occurrence in region $r_2$. The quadrants in Figure 1 were chosen by a principal components analysis of the February 1932- December 2004 mainshock M $\ge 4.0$ earthquakes in the region. The line between East and West corresponds roughly to the San Andreas Fault.

For the state-specific observation distributions,  each observation $y$ is a vector $(u,v)$, where $u>0$ is the interevent time in days since the previous earthquake and $v \in \{1,2\}$ is the region  in which the event occurred.  Given a state $s \in \mathcal{S}$, we assume an exponential distribution for interevent time, with mean $\lambda_s$, and a probability vector (depending on $s$) for the regions. We assume that, conditional on the state, interevent time and location are independent. Thus the state-specific densities are: for  $s \in \mathcal{S}$ and $y=(u,v), $
 
 \begin{equation} 
 p_s(y)=p_s(u,v)=e_s(u)q_s(v)
 \end{equation}
 
 where
 
 \begin{equation}  e_s(u)= (1/\lambda_s)e^{-u/\lambda_s} 
 \end{equation} 
 
 and $q_s=(q_s(1),q_s(2))$, with $ q_s(v)=$ P(the earthquake occurs in region $v \, | \,$system is in state $s$). Conditional on the state, the interevent time location distributions are taken to be exponential. The observation variable $Y_t$ is thus replaced by the vector $(U_t,V_t)$ and the desired scheduled forecast density for the next earthquake of an additional interevent time $u$ in the region $v$, given the past observations and an observed time of $w$ days since the most recent earthquake $P(U_{t+1}-w=u, V_{t+1}=v \, | \, y_1, \ldots, y_t, U_{t+1} \ge w).$
 
 As shown in the appendix, this is given by:

\begin{equation}
P(U_{t+1}-w=u, V_{t+1}=v \, | \, y_1, \ldots, y_t, U_{t+1} \ge w) = \sum_{s \in \mathcal{S}}d_{st}(w) (1/\lambda_s)e^{-u/\lambda_s}q_s(v) \\
\end{equation}
where 
\begin{equation}
d_{st}(w)=\frac{c_{st}e^{-w/\lambda_s}}{\sum_{r \in \mathcal{S}}c_{rt}e^{-w/\lambda_r}}.
\end{equation}

Paralleling (15), the scheduled forecast probability of an earthquake within $N$ days in quadrant $v$ is obtained by taking the $t$ interevent times for the earthquakes that have occurred by that time and computing the elapsed time $w$ since the most recent earthquake and then integrating (22) with respect to $u$ from $0$ to $N$, to give:

\begin{equation}
P(U_{t+1}-w \le N, V_{t+1}=v \, | \, y_1, \ldots, y_t, U_{t+1} \ge w) = \sum_{s \in \mathcal{S}}d_{st}(w) (1-e^{-N/\lambda_s})q_s(v) \\
\end{equation}

For the East/West model with (relabeled) state space (1,2,3,4)=((1,1),(2,1),(1,2),(2,2)) , the HMM parameters were estimated by:
\begin{equation}
\hat \pi=(0,0,1,0)
\end{equation}

\begin{equation}
(\hat \lambda_1,\hat \lambda_2,\hat \lambda_3,\hat \lambda_4)=(2.02,  21.59, 5.12, 22.82)
\end{equation}

\begin{equation}
\hat A=\begin{pmatrix}.512 & .475&.013&0 \cr
.041&0& .372&.587\cr
.032 & .031 & .625 & .311 \cr
.005 & .117 & .733 & .145 \cr
\end{pmatrix}
\end{equation}

\begin{equation}
\begin{pmatrix}\hat q_1 \cr \hat q_2 \cr \hat q_3 \cr \hat q_4  \cr \end{pmatrix}= \begin{pmatrix} 
1&0\cr
.88&.12\cr
0&1\cr
.08&.92\cr

\end{pmatrix}
\end{equation}

Thus, in our model, the system starts in the state corresponding to (a tendency for) short interevent time and location in the West; the mean interevent time given a tendency for short (long) interevent time and location in the East is 2.02 days (21.59 days); for the West, these are 5.12 and 22.82 days.  In the Markov chain, it's impossible to go from state (1,1) (short time, East location) to  state (2,2) (long time, West) or to go from state (2,2) to (2,2) in one step. For the states corresponding to short interevent times, the location distributions for East and West are estimated to put probability one on those regions; the probabilities for this are less than one given a long interevent time state.

Beginning on June 16, 1982, we made daily forecasts of the probability of an earthquake within one day (and within ten days) in each of the East/West locations, based on the data available at the forecast time; these data consisted of the interevent times and East/West locations of the earthquakes seen so far and the elapsed time $w$ since the most recent earthquake. (For the observation history for the first forecast, we again used the data  from the previous 30 earthquakes.) The desired forecast probabilities  followed immediately from (23) with $N=1$ day (and $N=10$), where $d_{st}(w)$ was calculated from (22), (10), and (3). This was repeated daily until December 28, 2008, resulting in 9693 forecasts.

To evaluate the time and location forecasts, we examined the two regions separately. For each of the regions, we split the forecasts into two groups, 9000 in a low range and 693 in a high range.  Within each group, we then found the proportion of days on which an earthquake did occur in that region within 24 hours (and within ten days) and compared it to the interval, mean, and median forecast probabilities for that group. The one day results are given in Table 4 and Figure 8 and the ten day results in  Figure 9.

[INSERT TABLE 4 ABOUT HERE]

[INSERT FIGURE 8 ABOUT HERE]

[INSERT FIGURE 9 ABOUT HERE] 

 The fit between one day  forecasts and observations in each group (Figure 8) appears quite good, similar to those for the entire study region (Figure 4). Similarly, the ten day forecasts for the East and West regions (Figure 9) also appear to fit the observations well.

In summary, the East/West model incorporated location as well as time in the forecasts. It performed well, generating forecasts that were close to the observations.

 \section{Discussion}
 
 A hidden Markov model consists of a sequence of observations and a sequence of hidden states.   In this application, a state is a statistical construct, not a physical one, but it corresponds to the idea that physical conditions in a given region yield a tendency for a mainshock earthquake to occur with shorter or longer interevent times. The key to our forecasting method is that, at each forecasting time, we can use the data available at that time to estimate the probability of being in each state and combine this with the state specific distributions to find and utilize the probability density of the next as-yet-unseen observation. 
  
We developed a two-state HMM and estimated the parameters using a selected portion of an earthquake catalog for southern California and western Nevada. Using this model we then issued, retroactively, a series of forecast probabilities of a mainshock earthquake occurring anywhere in the study region within one, five,  and ten days, each forecast being generated using the data known at that time. The forecasts were close to the observed seismicity.

We then extended the model and forecasts to include location  as well as time in an East/West model. In this model, the four states corresponded to a tendency for a shorter or longer interevent time in each of the East and West regions. We forecast probabilities of an earthquake occurring within one and ten days in each of the regions and compared these to what occurred. The forecasts were again  close to the observed seismicity.

It is important to recognize the real challenges in earthquake forecasting and limitations in our method. For example, we formulated a second four state HMM, called the North/South model. In this model, we combined quadrants 2 and 3 to form the North region and quadrants 1 and 4 to form the South region (Figure 1). We estimated the parameters using the same training set (1932-1964) used for the East/West model and computed forecasts for this period. These forecasts did not reflect the observed seismicity for this period; since the model didn't fit the training set, we did not proceed further.

There are several possible reasons why the North/South model failed while the East/West model succeeded. There are important geological and tectonic differences between the East/West and North/South divisions of the study area. The former model splits the study area along a fault line that delineates a plate boundary, while the latter doesn't. There is, therefore, no reason to suspect that a model fitting one pair of zones well should imply a model fitting the other pair will be successful. 

Another important limitation in our procedure is that its successful use relies on stationarity of the seismicity- that is, statistical similarity during the forecasting period to that of the time used in estimating the parameters. As seen in Figure 2, the distribution of interevent time in, roughly, years 1964-1982 appeared different than that of the rest of the 1932-2008 catalog; specifically, there were many more long times between earthquakes. 

 We generated forecasts for the period 1964-2004, which overlapped the nonstationary period. These forecasts were unsuccessful. Both the low and high range of forecasts overestimated the probability of an earthquake in a given time period compared to the observed proportion. Given the increased appearance of long interevent times, this is not surprising, but it highlights a serious limitation in fthe use of past seismicity to forecast characteristics of future seismicity. If the statistical behavior of seismicity changes, forecasts using a model based on an earlier period is not likely to be reliable. 

Finally, there is the challenge of estimating the HMM parameters on the basis of limited data. We used the first half of the catalog, consisting of 601 earthquakes from 1932-1964, to estimate the five parameters in the two state model and the 23 parameters in the East/West model. In general, training sets are much larger. 

In summary, limitations of the procedure are that forecasts appear to be reliable only for relatively large carefully chosen regions and, not surprisingly, good behavior of the forecasts depends on stationarity of the seismicity. To obtain good estimates of the HMM, it is necessary to have sufficient data. Within these admittedly important limitations, the HMM forecast methodology appears to be successful in identifying times in which there is an increased likelihood of a mainshock in a given time interval and region.

\appendix
\section{Appendix}

In this appendix, we present the well-known parameter estimation procedure for hidden Markov models,  derive the post-event and scheduled forecast densities, and give a rigorous proof of the waiting time paradox result.

\subsection{Estimating parameters}
Recall the forward and backward variables $f_s(t)$ and $b_s(t)$ can be computed using (3) and (4)
when the parameters of the HMM are known and the complete set of observations $\mathcal{O}$ is seen. These variables can then be used to find  $\gamma_s(t) \equiv P(X_t=s | \, \mathcal{O})$, the probability of being in state $s \in \mathcal{S}$ at time $t $, conditional on all the observations. By conditioning on the event $(X_t=s, y_1, \ldots, y_t)$ and using (2), it is easily seen that
\begin{equation}
\gamma_s(t)=b_s(t)f_s(t)/P(\mathcal{O}), ~1 \le t \le L.
\end{equation}
Similarly,  we can compute $\eta_{rs}(t) \equiv P(X_t=r, \, X_{t+1}=s |  \mathcal{O})$, the joint probability of being in state $r$ at time $t$ and in state $s$ at time $t+1$, conditional on the complete set of observations. By definition,
\begin{equation}
\eta_{rs}(t)= P(X_t=r, X_{t+1}=s, y_1 \ldots, y_L)/P(\mathcal{O})
\end{equation}
The denominator in (29) is given by (6).
By a pair of conditioning arguments and simplification by the conditional independence in (2), we have:
\begin{equation}
\eta_{rs}(t) = P(X_t=r, \, X_{t+1}=s | \mathcal{O} )=f_r(t)a_{rs}b_s(t+1)p_s(y_{t+1})/P(\mathcal{O}).
\end{equation}

In summary, given the complete set of observations $\mathcal{O}$ and set of parameters $\Theta$, we can compute $f_s(t), \, b_s(t), \, \gamma_s(t)$, and $ \eta_{rs}(t)$, for all states $r, s \in S$ and times $t=1, \ldots L$.

To estimate the parameters of a HMM, we use the well-known Baum Welch algorithm (Rabiner, 1989), described below. In the following, we replace $P$ by $P_\Theta$ to emphasize the dependence on the parameter set.

The procedure begins with an initial guess for the parameters, say $\Theta^{(0)}$. Then $f_s^{(0)}(t), \, b_s^{(0)}(t), \, \gamma_s^{(0)}(t)$, and $ \eta_{rs}^{(0)}(t)$ $(r, s \in S$, $t=1, \ldots, L$), and $P_\Theta^{(0)}(\mathcal{O} )$ are computed using (3)-(6), (28), and (30).
We then update the parameters as below to get a new estimate $\Theta^{(1)}$. The key is that the likelihood of the observations increases under the new parameter set: $P_\Theta^{(1)}(\mathcal{O} ) \ge P_\Theta^{(0)}(\mathcal{O} )$ (Rabiner, 1989). The procedure is then iterated until the likelihoods converge to some desired precision or until successive parameter estimates differ by a predetermined small amount. The parameters are estimated by this final updating.

The parameter updates are:
\begin{equation}
a_{rs}^{(n+1)}= \frac{\sum_{t=1}^{L-1} \eta_{rs}^{(n)}(t)}{\sum_{t=1}^{L-1}\gamma_r^{(n)}(t)}
\end{equation}
\begin{equation}
\pi^{(n+1)}_s= \gamma_s^{(n)}(1)
\end{equation}
\begin{equation}
\lambda_s^{(n+1)}= \frac{\sum_{t=1}^L \gamma_s^{(n)}(t)y_t}{\sum_{t=1}^L \gamma_s^{(n)}(t)}
\end{equation}
\begin{equation}
q_s^{(n+1)}(v)=\frac{\sum_{t=1, v_t=v}^L \gamma_s^{(n)}(t)}{\sum_{t=1}^L \gamma_s^{(n)}(t)}
\end{equation}

where $y_t$ is the $t^{th}$ observation.

The transition probability estimate in (31) is  simply the ratio of the expected number of transitions from  $r$ to $s$  to the expected number of transitions from $r$. The initial distribution  in (32) is the probability of starting in state $s$ under $\Theta^{(n)}.$ For a single exponential density, the maximum likelihood estimator of the mean is the sample mean; here, for state $s$, (33) weights each observation by the probability it came from  state $s.$  In (34), the probability the event occurs in region $v$ given the state is $s$ is estimated by the ratio of the expected number of times this occurs to the expected number of visits to the state.

Finally, note that estimation procedures exist for other state-specific distributions, including  normal distributions (Rabiner, 1989). Obvious extensions to the forecast formulas in Section 3 are then possible. 

\subsection{Forecast densities}

We first derive the (post-event) density function for waiting time until the next earthquake, made immediately after an earthquake. We first derive this density in the framework of a general HMM. The problem is to find the probability density function of observation $Y_{t+1}$ conditional on the observation history up to time $t$- that is, $P(Y_{t+1}=y \, | \, y_1, \ldots,y_t)$.  By summing over all possible states at time $t+1$, standard conditioning arguments, and (1), 
$$P(Y_{ t+1}=y \, | \, y_1, \ldots ,y_ t)= \sum_{s \in \mathcal{S}}  P(Y_{ t+1}=y \, | \, X_{ t+1}= s, y_1, \ldots , y_ t)P(X_{ t+1}= s \, | \,  y_1, \ldots ,y_ t)=$$
$$\sum_{s \in \mathcal{S}}  \sum_{r \in \mathcal{S}}   P(Y_{ t+1}=y \, | \, X_{ t+1}= s)P(X_{ t+1}= s \, | \,   X_ t=r, y_1, \ldots ,y_ t) \times $$
$$ P( X_ t=r  \, | \,   y_1, \ldots ,y_ t)=$$
$$ \sum_{s \in \mathcal{S}}  \sum_{r \in \mathcal{S}}  p_ s(y) P(X_{ t+1}= s \, | \,   X_ t=r) P( X_ t=r ,  y_1, \ldots ,y_ t  )/P(y_1, \ldots ,y_ t )$$
$$ =\sum_{s \in \mathcal{S}}  \sum_{r \in \mathcal{S}}  p_ s(y) a_{r s} f_r( t)/P(y_1, \ldots ,y_ t ).$$

That is, at time $t$, the distribution of the next observation, given the observations available, is computable by
\begin{equation}
P(Y_{t+1}=y | y_1, \ldots , y_t)= \sum_{s \in \mathcal{S}} c_{st}p_s(y) 
\end{equation}
where 
\begin{equation}
c_{st}=\sum_{r \in \mathcal{S}}a_{rs} f_r(t)/P(y_1, \ldots y_t).
\end{equation}
Note that $\sum_{s \in \mathcal{S}}c_{st}=\sum_{r \in \mathcal{S}} f_r(t)(\sum_{s \in \mathcal{S}} a_{rs})/P(y_1, \ldots, y_t )$
$=\sum_{r \in \mathcal{S}}f_r(t)/P(y_1, \ldots, y_t )=1$ by (5), hence
\begin{equation}
\sum_{s \in \mathcal{S}}c_{st}=1
\end{equation}
Furthermore, an inspection of the above shows that 
\begin{equation}
c_{st}= P(X_{t+1}=s \, | \, y_1, \ldots , y_t). 
\end{equation}
Thus, (35)-(38) show that the forecast density of the next observation, conditional on the data available at that time is a mixture of the state-specific densities, in which the weight for $p_s(y)$ is (a) calculable and (b) interpretable as the conditional probability the system is in that state $s$ given the data available at the forecast time.

Recall that in our application the observation $Y_t$ corresponding to earthquake $t$ is  the time in days since earthquake $t-1.$ Then, (7) and (35) give the density of the waiting time until the next earthquake, given the seismic history available just after an earthquake:
\begin{equation}
P(Y_{t+1}=y \, | \, y_1, \ldots y_t)=\sum_{s \in \mathcal{S}}c_{st}(1/\lambda_s)e^{-y/\lambda_s}.
\end{equation}

Next, we  extend our method to scheduled forecasts, say weekly or daily, of the probability of an earthquake within a given time period in the region. Now what is known at the time of the forecast is the set of observed interevent times as well as the elapsed time since the previous earthquake. Here we derive the forecast density given in  (13) and (14).

Suppose that $t$ earthquakes have occurred at the forecast time and $w$ days have elapsed since the most recent earthquake. The available data at forecast time is then  $y_1, \ldots , y_t$ and the event $Y_{t+1} \ge w$. The remaining waiting time until the next is $Y_{t+1}-w$. We derive its density. Let $y>0.$ The forecast density of the remaining waiting time, given the available data, is 
$$
P(Y_{t+1}-w=y \, | \, y_1, \ldots, y_t, Y_{t+1} \ge w)= \frac{P(Y_{t+1}=y+w, Y_{t+1}\ge w   \, | \, y_1, \ldots, y_t)}{P(Y_{t+1} \ge w \, | \, y_1, \ldots , y_t)}
$$
$$
=\frac{P(Y_{t+1}=y+w \, | \, y_1, \ldots, y_t)}{ P(Y_{t+1} \ge w \, | \, y_1, \ldots , y_t)}= \frac{P(Y_{t+1}=y+w \, | \, y_1, \ldots , y_t)}{\int_w^\infty P(Y_{t+1}=x \, | \, y_1, \ldots, y_t)dx}.
$$

The numerator and denominator in the final fraction, by (40), are 
$$
\mbox{$\sum_{s \in \mathcal{S}} c_{st}e^{-w/\lambda_s} (1/\lambda_s)e^{-y/\lambda_s}$ and $\int_w^\infty \sum_{r \in \mathcal{S}} c_{rt} (1/\lambda_r)e^{-x/\lambda_r} \, dx$,}
$$
respectively. The denominator simplifies to $
\sum_{r \in \mathcal{S}} c_{rt}e^{-w/\lambda_r}.$

That is, at the time of the forecast, if $t$ earthquakes have occurred and $w$ days have elapsed since the previous earthquake, the density of the remaining waiting time $Y_{t+1}-w$ is
\begin{equation}
P(Y_{t+1}-w=y \, | \, y_1, \ldots, y_t, Y_{t+1} \ge w) = \sum_{s \in \mathcal{S}}d_{st}(w) (1/\lambda_s)e^{-y/\lambda_s} \\
\end{equation}
where 
\begin{equation}
d_{st}(w)=\frac{c_{st}e^{-w/\lambda_s}}{\sum_{r \in \mathcal{S}}c_{rt}e^{-w/\lambda_r}}.
\end{equation}

It is easily shown that $d_{st}(w)=P(X_{t+1}=s \, | \, y_1, \ldots, y_t, Y_{t+1} \ge w)$, so that the forecast density is again a mixture of the densities $p_s(y)$ with the weight for  $p_s(y)$ being the conditional probability of being in state $s$ at the next earthquake, given all the information available at the forecast time.

The extension of the HMM and forecasts to include location as well as interevent time is straightforward. As mentioned in section 5, the observation $y$ is now a vector $(u,v)$, where $u>0$ is the interevent time in days and $v$ is the region where the earthquake occurred. The observation variable $Y_t=(U_t,V_t)$. The state specific density in (7) is replaced by $p_s(y)=p_s(u,v)= (1/\lambda_s)e^{-u/\lambda_s} q_s(v)$. By (35), the post-event forecast density in (39) becomes
 
\begin{equation}
P(U_{t+1}=u, V_{t+1}=v \, | \, y_1, \ldots, y_t)= \sum_{s \in \mathcal{S}}c_{st}(1/\lambda_s)e^{-u/\lambda_s}q_s(v)
\end{equation}

and the scheduled forecast density in (40) becomes 
\begin{equation}
P(U_{t+1}-w=u, V_{t+1}=v \, | \, y_1, \ldots, y_t, U_{t+1} \ge w) = \sum_{s \in \mathcal{S}}d_{st}(w) (1/\lambda_s)e^{-u/\lambda_s}q_s(v). \\
\end{equation}

\subsection{Waiting time paradox}

In section 3.4, we gave a heuristic argument that an HMM model for seismicity with state-specific exponential distributions satisfies the waiting time paradox; here is a rigorous proof.

Denoting $E(Y_{t+1}-w \, | \, y_1, \ldots y_t, Y_{t+1} \ge w)$ by $h(w)$, it follows from (16) and (14) that  
\begin{equation}
h'(w)= \sum_{s \in \mathcal{S}} \lambda_s \frac{d}{dw} \frac{c_{st}e^{-w/\lambda_s}}{\sum_{r \in \mathcal{S}}c_{rt}e^{-w/\lambda_r}}
\end{equation}
After differentiating and simplifying, this gives
\begin{equation}
h'(w)=\sum_{s \in \mathcal{S}}\sum_{r \in \mathcal{S}}c_{rt}e^{-w/\lambda_r}c_{st}e^{-w/\lambda_s}(\frac{\lambda_s}{\lambda_r} -1)/(\sum_{r \in \mathcal{S}} c_{rt}e^{-w/\lambda_r})^2
\end{equation}
The denominator is positive, so the sign of $h'(w)$ is determined by the numerator. Impose an order on the state space $\mathcal{S}$; since the terms vanish when $r=s$, the numerator in (45) can be rewritten as 
\begin{equation}
\sum_{s \in\mathcal{S}}\sum_{r \le s}c_{rt}e^{-w/\lambda_r}c_{st}e^{-w/\lambda_s}(\frac{\lambda_s}{\lambda_r}-1)+
\sum_{s \in\mathcal{S}}\sum_{r \ge s}c_{rt}e^{-w/\lambda_r}c_{st}e^{-w/\lambda_s}(\frac{\lambda_s}{\lambda_r}-1)
\end{equation}
In the second sum reverse the order of summation and then make a change of summation variables $(r \leftrightarrow s)$ to see that (47) equals
\begin{equation}
\sum_{s \in\mathcal{S}}\sum_{r \le s}c_{rt}e^{-w/\lambda_r}c_{st}e^{-w/\lambda_s}(\frac{\lambda_s}{\lambda_r}-1+ \frac{\lambda_r}{\lambda_s}-1)
\end{equation}
Since $(\lambda_s/\lambda_r-1+ \lambda_r/\lambda_s-1)=(\lambda_s-\lambda_r)^2/(\lambda_r \lambda_s) > 0$, (45)-(48) show $h'(w) > 0$ for all $w>0$ and thus $E(Y_{t+1}-w| y_1, \ldots y_t, Y_{t+1} > w)$ is an increasing function of $w$ and our model satisfies the waiting time paradox.

\subsection{Acknowledgements}
The authors thank two anonymous reviewers for their helpful comments and suggestions.
\newpage

\begin{center}
References
\end{center}

Baum, L. E. and Petrie, T. (1966). Statistical inference for probabilistic functions of finite state Markov chains, {\it Annals of Mathematical Statistics}, 37:1554-1563.  

Baum, L.E., Petrie, T., Soules, G., and Weiss, N. (1970). A maximization technique in the statistical analysis of probabilistic functions of Markov chains, {\it The Annals of Mathematical Statistics}, 41, 164-171.

Baum, L.E. (1972). An inequality and associated maximization technique in statistical estimation for probabilistic functions of Markov chains, {\it Inequalities}, 3, 1-8.

Beyreuther, M., Carniel, R., and Wasserman, J. (2008). Continuous hidden Markov models: application to automatic earthquake detection and classification at Las Can${\tilde{\rm  a}}$das caldera, Tenerife, {\it Journal of Volcanology and Geothermal Research} 176, 513-518.

Chambers, D.W., Ebel, J.E., Kafka, A.L., and Baglivo, J.A. (2003). Hidden Markov approach to modeling interevent earthquake times, {\it Eos. Trans. AGU}, 84 (46), Fall Meet. Suppl., Abstract S52F-0179.
 
Clote, P. and R. Backofen (2000). Computational Molecular Biology, An Introduction. New York: John Wiley and Sons.
  
Davis, P.M., Jackson, D.D., and Kagan, Y.Y. (1989). The longer it has been since the last earthquake, the longer the expected time till the next? {\it Bull. Seism. Soc. Am.} 79 (5), 1439-1456.
 
Durbin, R.,  Eddy, S.,  Krogh, A.,  and  Mitchison, G. (1998). Biological Sequence Analysis: Probabilistic Models of Proteins and Nucleic Acids. Cambridge: Cambridge University Press.
  
Ebel, J.E, Chambers, D.W., Kafka, A.L., and Baglivo, J.A (2007). Non-Poissonian earthquake clustering and the hidden Markov model as bases for earthquake forecasting in California, {\it Seismological Res. Lett.,} 78 (1), 57-65.
    
Fredkin, D.R. and Rice, J.A. (1992a). Bayesian restoration of single-channel patch clamp recordings, {\it Biometrics}, 48 (2), 427-448.

Fredkin, D.R. and  Rice J.A. (1992b). Maximum likelihood estimation and identification directly from single-channel recordings, {\it Proceedings: Biological Sciences}, 249, 125-132.

Gardner, J.K. and Knopoff, L. (1974). Is the sequence of earthquakes in southern California with aftershocks removed Poissonian? {\it Bull. Seism. Soc. Am.} 64, 1363-1367.
  
Granat, R. and Donnellan, A. (2002). A hidden Markov model based tool for geophysical data exploration, {\it Pure and Applied Geophysics}, 159, 2271-2283.
 
Rabiner, L.R. (1989). A tutorial on hidden Markov models and selected applications in speech recognition, {\it Proc. IEEE} 77, 257-286.
 
Sornette, D. and Knopoff, L. (1997). The paradox of the expected time until the next earthquake, {\it Bull. Seism. Soc. Am.} 87 (4), 789-798.

Tsapanos, T.M. (2001). The Markov model as a pattern for earthquakes recurrence in South America, {\it Bull. Geol. Soc. Greece}, 39 (4), 1611-1617.

Wu, Z. (2010). A hidden Markov model for earthquake declustering, {\it J. Geophys. Res.}, 115, B03306, doi:10.1029/2009JB005997
\vfill
\eject

\begin{figure}[h]
\begin{center}
\includegraphics{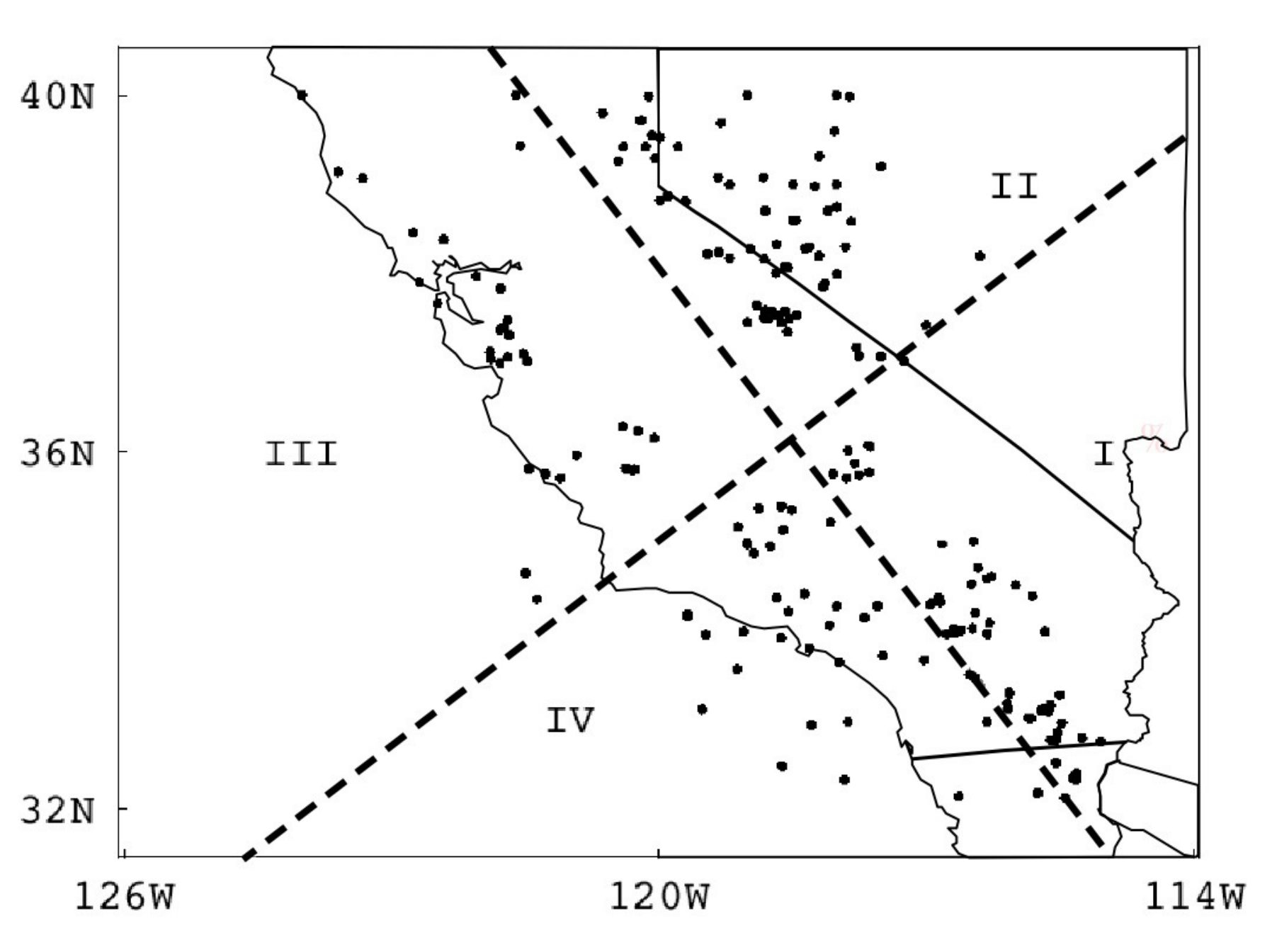}
\end{center}
\caption{\it Study region with four quadrants defined using principal components axes of event locations between February 1932 and December 2004.}
\label{fig1}
\end{figure}

\newpage

\begin{figure}[h]
\begin{center}
\includegraphics{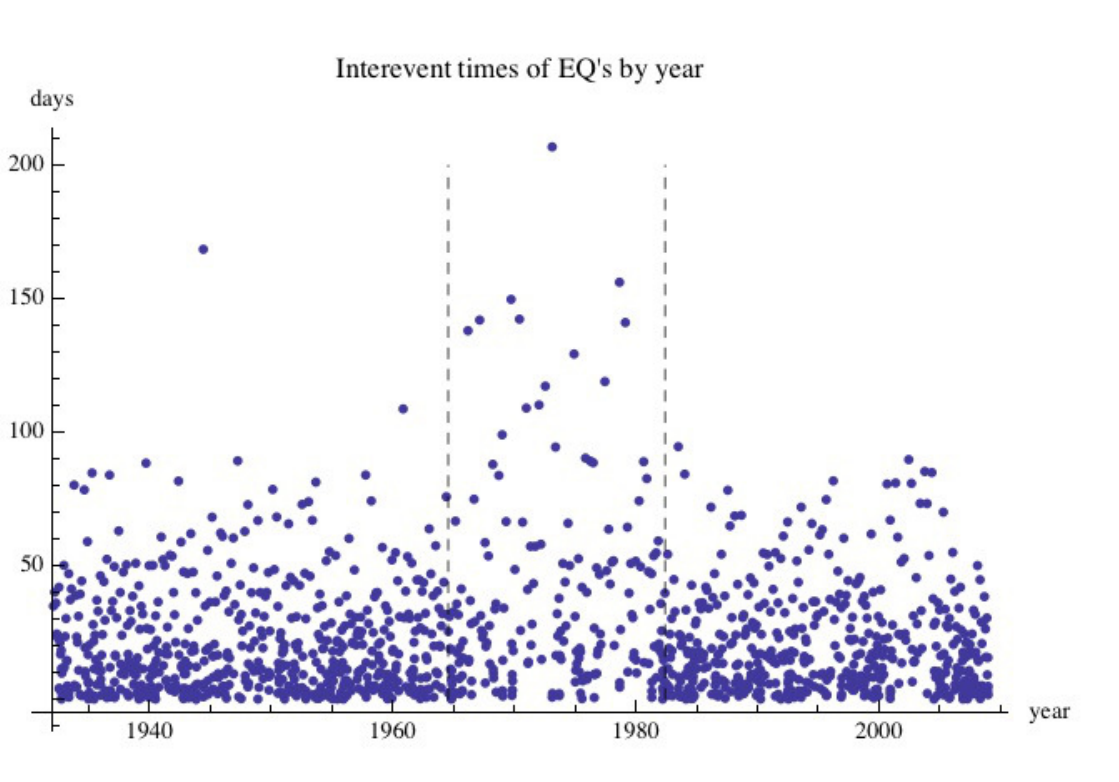}
\end{center}
\caption{\it Days between earthquakes (vertical axis) versus year (horizontal axis). Vertical dashed lines correspond to July 1964 and June 1982.}
\label{fig2}
\end{figure}

\newpage

\begin{figure}[h]
\begin{center}
\includegraphics{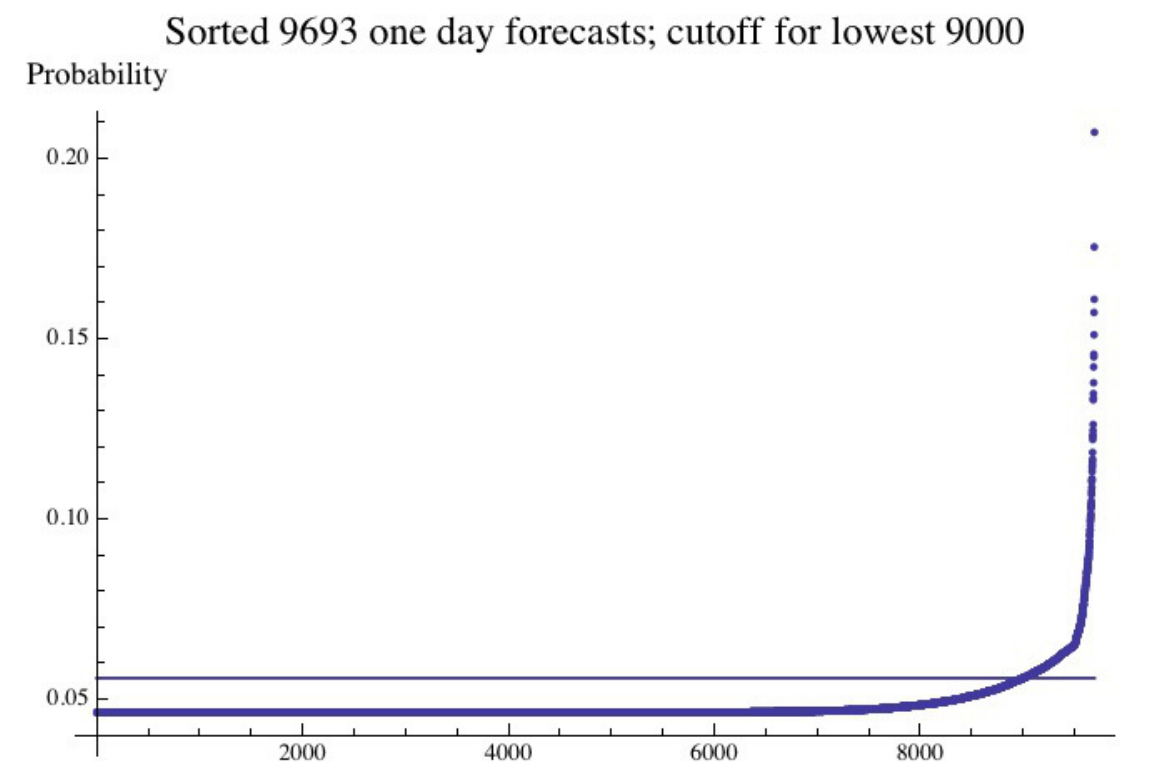}
\end{center}
\caption{\it Sorted one day forecast (vertical axis) versus forecast number (horizontal axis).  Horizontal line separates forecasts into groups of 9000 and 693.}
\label{fig3}
\end{figure}

\newpage

\begin{figure}[h]
\begin{center}
\includegraphics{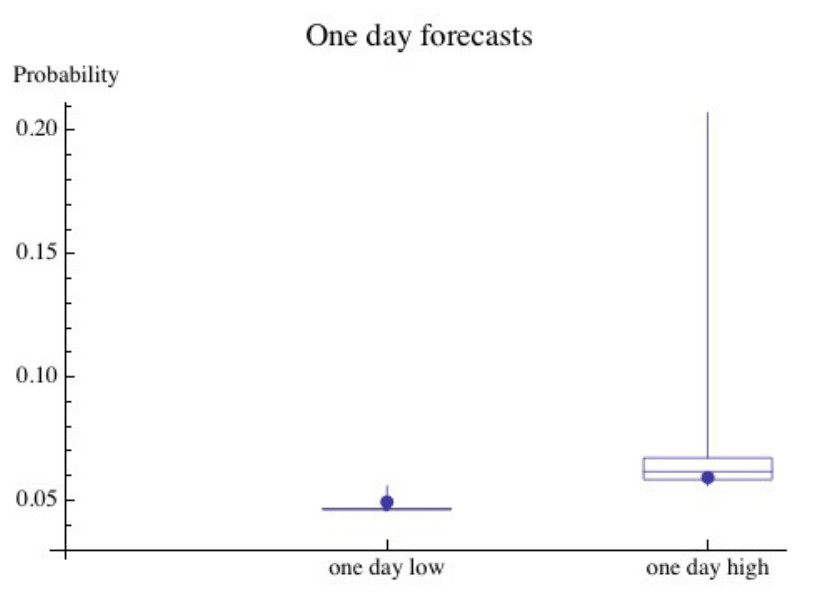}
\end{center}
\caption{\it 9693 daily forecast probabilities of an earthquake in one day, divided into 9000 lowest and 693 highest forecasts. Lines give range of values and boxes give first quartile, median, and third quartile of forecasts in that range. Circles are observed proportion of times an earthquake occurred within one day among those forecasts.}
\label{fig4}
\end{figure}

\newpage

\begin{table}[h]
\begin{center}
\renewcommand{\arraystretch}{1.25}
{\small\begin{tabular}{|ccccc|cc|}
\hline
& \multicolumn{4}{c|}{One Day Forecasts}& \multicolumn{2}{c|}{Observations}\\
&range&number &mean&median&number &proportion\\
\hline
low&(.0463,\,.0558)&9000&.0470&.0463&444&.0493\\
high&(.0558,\,.2072)&693&.0672&.0616& 41&.0592\\
\hline
\end{tabular}}
\caption{\it 9693 daily forecast probabilities of an earthquake in one day, divided into two groups; for each group observed proportion of times this occurred.}
\end{center}
\end{table}

\newpage

\begin{figure}[h]
\begin{center}
\includegraphics{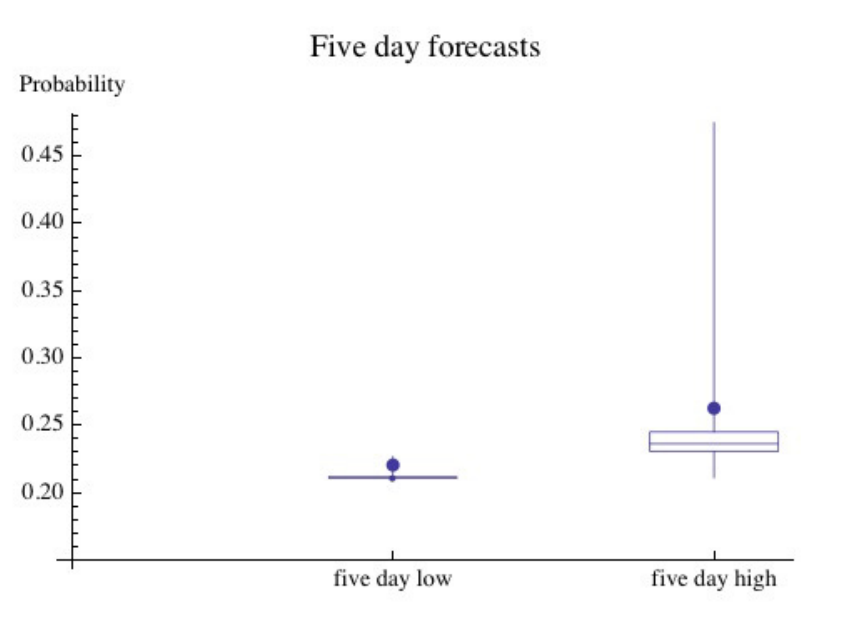}
\end{center}
\caption{\it 9693 daily forecast probabilities of an earthquake in five days, divided into 9000 lowest and 693 highest forecasts. Lines give range of values and boxes give first quartile, median, and third quartile of forecasts in that range. Circles are observed proportion of times an earthquake occurred within five days among those forecasts.}
\label{fig5}
\end{figure}

\

\

\begin{table}[h]
\begin{center}
\renewcommand{\arraystretch}{1.25}
{\small\begin{tabular}{|ccccc|cc|}
\hline
& \multicolumn{4}{c|}{Five Day Forecasts}& \multicolumn{2}{c|}{Observations}\\
&range&number &mean&median&number &proportion\\
\hline
low&(.2110,\,.2266)&9000&.2121&.2110&1983&.2203\\
high&(.2266,\,.4748)& 693&.2453&.2361& 182&.2626\\
\hline
\end{tabular}}
\caption{\it 9693 daily forecast probabilities of an earthquake in five days, divided into two groups; for each group observed proportion of times this occurred.}\end{center}
\end{table}

\newpage
\begin{figure}[h]
\begin{center}
\includegraphics{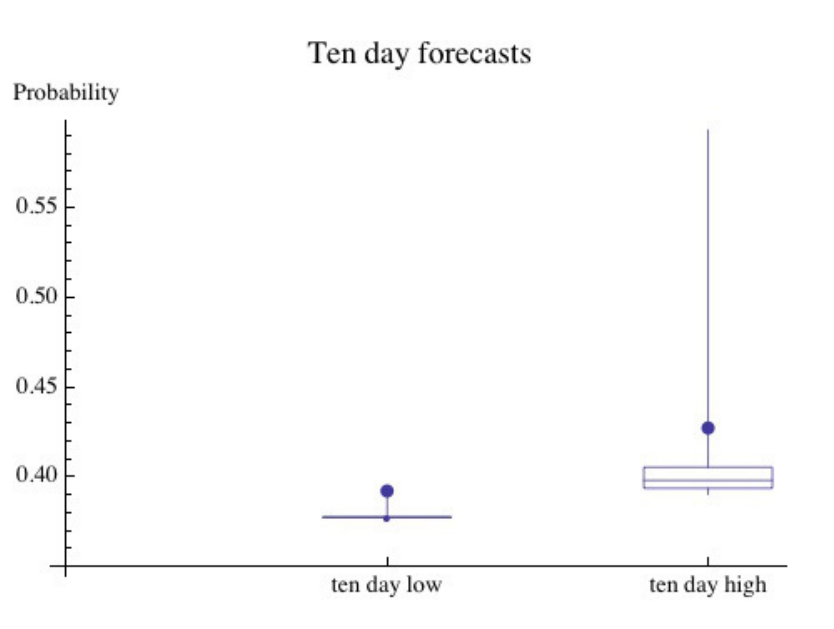}
\end{center}
\caption{\it 9693 daily forecast probabilities of an earthquake in ten days, divided into 9000 lowest and 693 highest forecasts. Lines give range of values and boxes give first quartile, median, and third quartile of forecasts in that range. Circles are observed proportion of times an earthquake occurred within ten days among those forecasts.}
\label{fig6}
\end{figure}

\

\

\begin{table}[h]
\begin{center}
\renewcommand{\arraystretch}{1.25}
{\small\begin{tabular}{|ccccc|cc|}
\hline
& \multicolumn{4}{c|}{Ten Day Forecasts}& \multicolumn{2}{c|}{Observations}\\
&range&number &mean&median&number &proportion\\
\hline
low&(.3775,\,.3902)&9000&.3784&.3775&3529&.3921\\
high&(.3902,\,.5930)& 693&.4055&.3980& 296&.4271\\
\hline
\end{tabular}}
\caption{\it 9693 daily forecast probabilities of an earthquake in ten days, divided into two groups; for each group observed proportion of times this occurred.}
\end{center}
\end{table}

\begin{figure}[h]
\begin{center}
\renewcommand{\arraystretch}{1.25}
\includegraphics{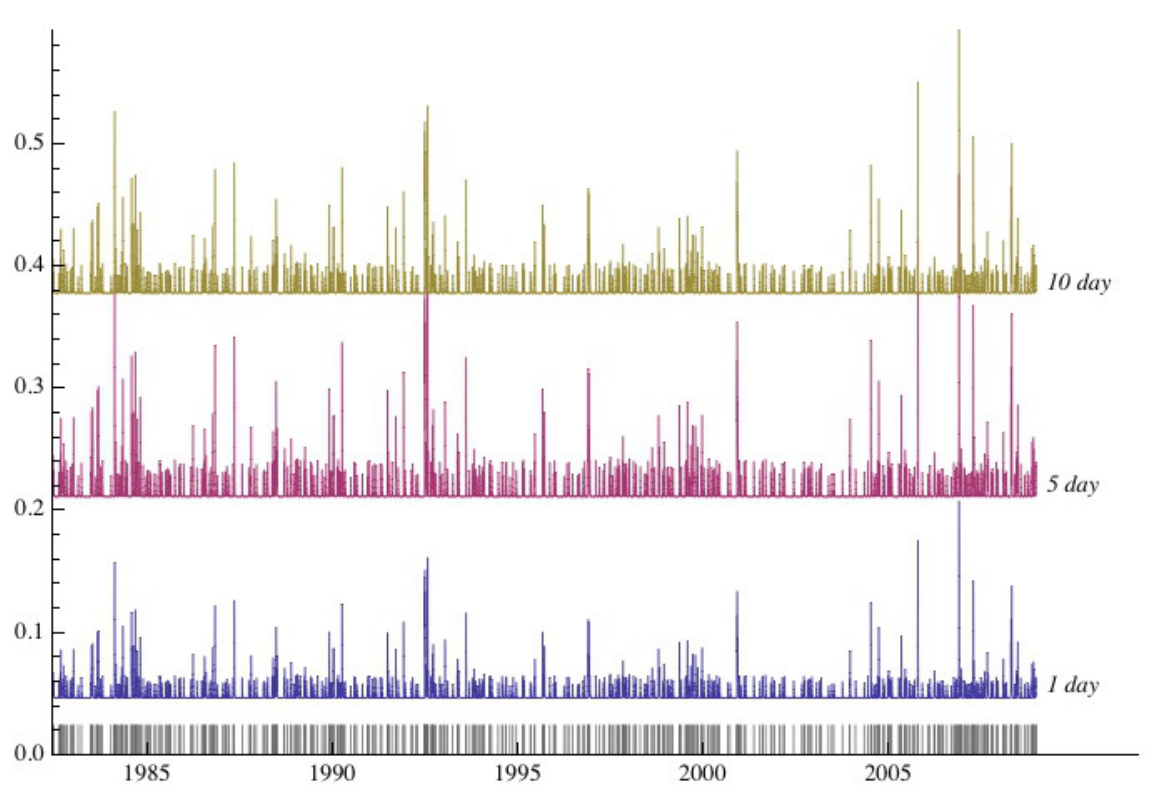}
\end{center}
\caption{\it One day, five day, and ten day forecasts for the period from June 1982 through December 2008, and dates of earthquake events.}
\label{fig7}
\end{figure}

 \begin{table}[h]
\begin{center}
\begin{tabular}{|c|ccccc|cc|}
\hline
 & & \multicolumn{4}{c|}{One Day Forecasts}& \multicolumn{2}{c|}{Observations}\\
  Region & & range & number & mean & median   &   number & 
  proportion\\
  \hline
   East &  low &(.0193,\, .0326)& 9000 &.0287 &  .0279 & 232& .0258\\
   & high & (.0326,\, .1722) & 693 &.0429& .0362 &   28 & .0404\\ 
 West & low & (.0124,\, .0255) & 9000 & .0182 & .0177 &  207 & .0230\\
  & high & (.0255,\, .1040) & 693 & .0342 & .0275 &  18 & .0260\\
  \hline
  \end{tabular}
  \caption{\it 9693 daily forecast probabilities of an earthquake in one day in each of the East/West regions; for each region forecasts divided into two groups; observed proportion of times an earthquake occurred in this region within one day.}
  \end{center}
  \end{table}

\

\

\begin{figure}[h]
\begin{center}
\includegraphics{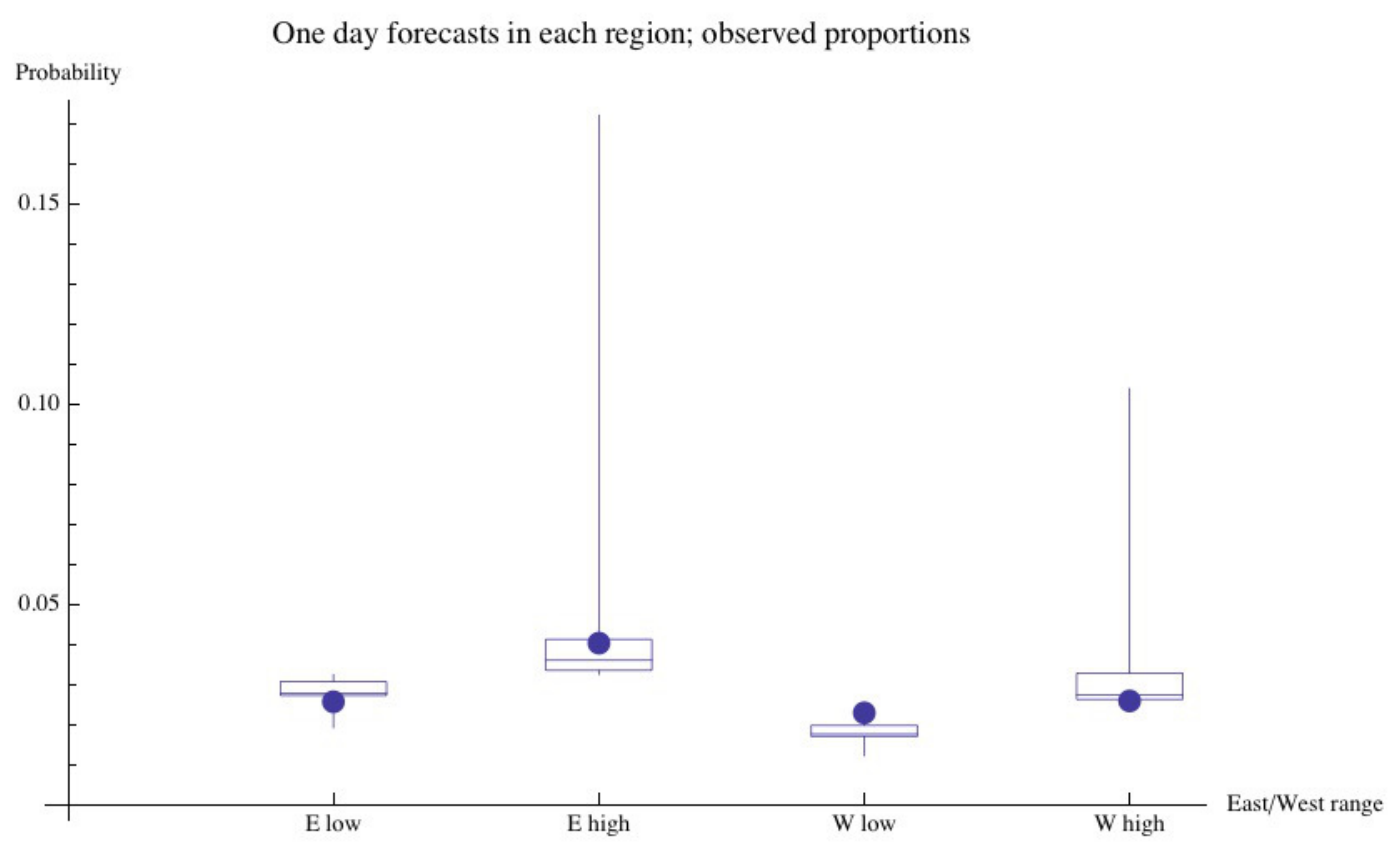}
\end{center}
\caption{\it 9693 daily forecast probabilities of an earthquake in one day in the East/West regions, each divided into  low and high ranges of forecasts. Boxes give first quartile, median, and third quartile of forecasts in that range. Circles are observed proportion of times an earthquake occurred in one day in that region among that range of forecasts.}
\label{fig8}
\end{figure}

\begin{figure}[h]
\begin{center}
\includegraphics{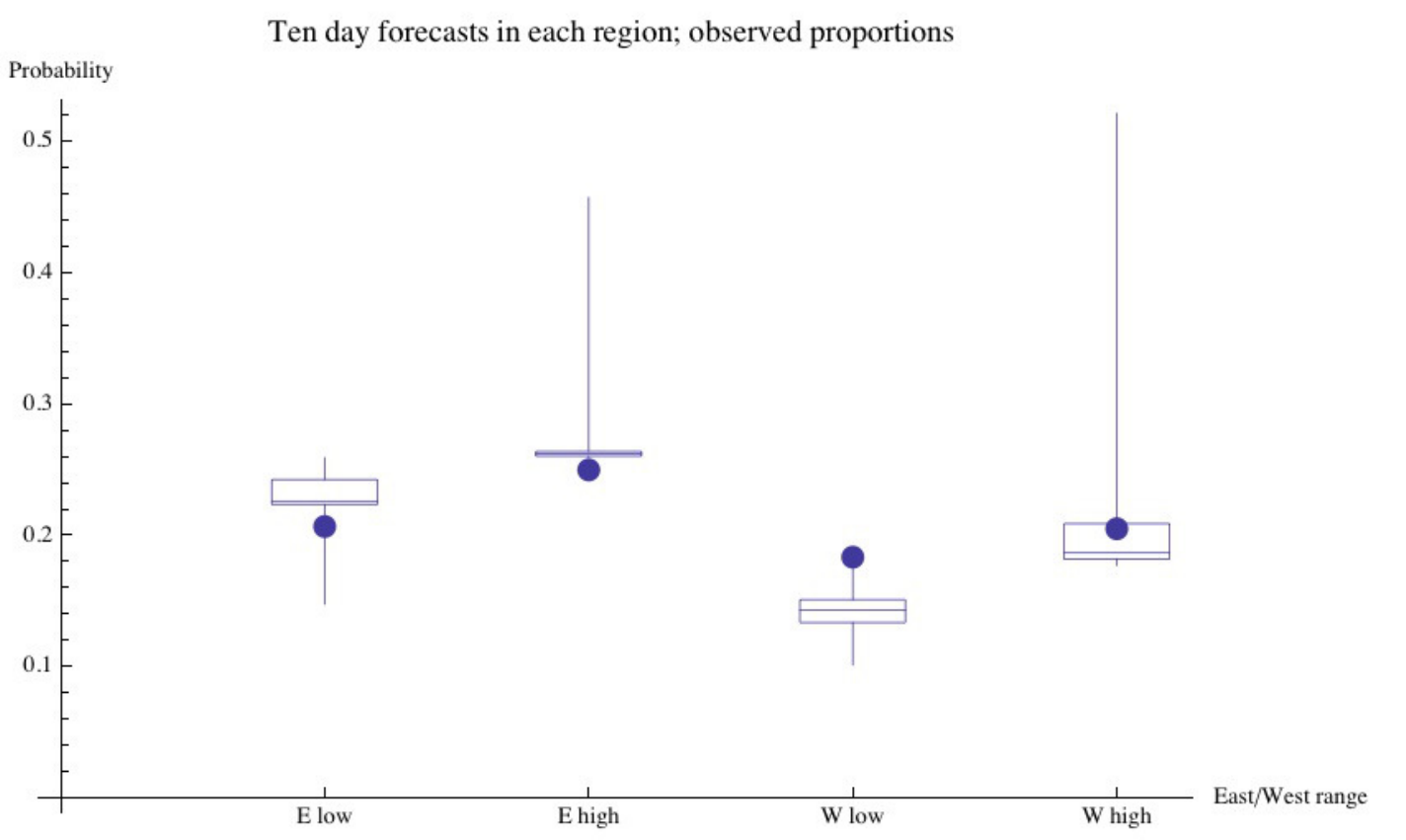}
\end{center}
\caption{\it 9693 daily forecast probabilities of an earthquake in ten days in the East/West regions, each divided into  low and high ranges of forecasts. Boxes give first quartile, median, and third quartile of forecasts in that range. Circles are observed proportion of times an earthquake occurred in ten days in that region among that range of forecasts.}
\label{fig9}
\end{figure}

\end{document}